\documentclass[%
 aip,  jmp, amsmath,amssymb,
reprint,%
]{revtex4-1}

\usepackage{graphicx}
\usepackage{dcolumn}
\usepackage{bm}

\begin{document}

\title{Spin echo dynamics under an applied drift field in graphene nanoribbon superlattices } 

\author{Sanjay Prabhakar}
\email[]{sprabhakar@wlu.ca}
\homepage[]{http://www.m2netlab.wlu.ca}
\affiliation{M\,$^2$NeT Laboratory, Wilfrid Laurier University, 75 University Avenue West, Waterloo, ON, Canada, N2L 3C5
}
\author{Roderick Melnik}
\affiliation{M\,$^2$NeT Laboratory, Wilfrid Laurier University, 75 University Avenue West, Waterloo, ON, Canada, N2L 3C5
}
\affiliation{
Gregorio Millan Institute, Universidad Carlos III de Madrid, 28911, Leganes, Spain
}
\author{Luis L Bonilla}
\affiliation{
Gregorio Millan Institute, Universidad Carlos III de Madrid, 28911, Leganes, Spain
}
\author{James E Raynolds}
\affiliation{
Drinker Biddle $\&$ Reath LLP, Washington DC 20005, USA.
}

\date{October 24, 2013}

\begin{abstract}
We investigate the evolution of spin dynamics in graphene nanoribbon superlattices (GNSLs) with armchair and zigzag edges in the presence of a drift field. We determine the exact evolution operator and show that it exhibits spin echo phenomena due to rapid oscillations of the quantum states along the ribbon. The evolution of the spin polarization  is accompanied by strong beating patterns. We also provide  detailed analysis of the band structure of GNSLs with armchair and zigzag edges.
\end{abstract}



\maketitle

Manipulation of electron spins using gate potentials in low dimensional semiconductor nanostructures is of interest, among other things, in that it provides a promising  approach for the practical realization of robust qubit operations.~\cite{trauzettel07,wang-das-sarma12,barnes12,wang12,szumniak13,ban12,prabhakar13,elzerman04,prabhakar10,shi12} In recent years, experimental and theoretical research has sought a better understanding of the underlying physics of electrostatically defined quantum dots formed in two-dimensional electron gases for applications to solid state based quantum computing.\cite{pryor06,flatte11,prabhakar09,prabhakar11,prabhakar12,prabhakar13a,amasha08} In these devices, the spin-orbit interaction gives rise to decoherence due to the coupling of the electron spins to lattice vibrations. Hyperfine interactions between electron and nuclear spins are also a factor in some systems. Much work has focused on III-V systems although Si quantum dots~\cite{yang12} are also of interest because of their relatively long decoherence times due to weak spin-orbit and hyperfine interactions.~\cite{shi12} In another promising approach, experimentalists have succeeded in fabricating and testing a low operation voltage organic field effect transistor using graphene as the gate electrode placed over a thin polymer gate dielectric layer.~\cite{song13} Graphene is promising because it exhibits extremely weak spin orbit coupling and hyperfine interactions.~\cite{min06,trauzettel07,recher09,recher10,allen12,fuchs12,krueckl12,stefano12,lih-king12,jung12}
In this paper, we present a theoretical investigation of  the spin echo phenomena in GNSLs under an externally applied drift field. We find that  the  spin echo is accompanied by a  strong beating pattern in the evolution of spin dynamics along the GNSLs with armchair and zigzag edges. We show that with a particular choice of the period and drift field, the spin polarization can be controlled to propagate on the surface of the Bloch sphere in a desired fashion.

\begin{figure}
\includegraphics[width=8.5cm,height=7cm]{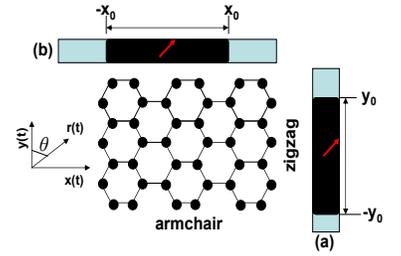}
\caption{\label{fig1}
(Color online) Schematic diagram of graphene sheet with armchair along x-axis and zigzag along y-axis. Spin orientation shown by arrow sign in Fig.~\ref{fig1} (a) and (b)    move rapidly between $-x_0$ and $+x_0$ for armchair and between $-y_0$ and $+y_0$ for zigzag GNSLs that induce spin echo under an externally applied drift field.
}
\end{figure}
\begin{figure}
\includegraphics[width=8.5cm,height=4cm]{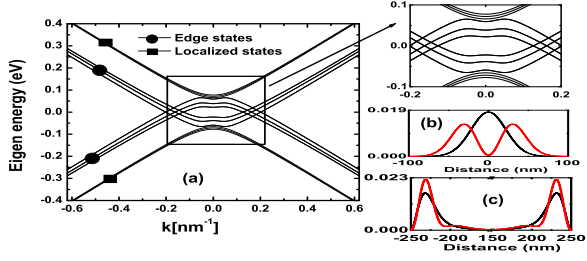}
\caption{\label{fig2}
(Color online) Band structure of GNSLs with zigzag edge. Localized and edge states wave function squared are plotted in Fig.~\ref{fig2} (b) and (c) at $k=0.1\mathrm{nm^{-1}}$.  Here we chose  $U_0=+60~\mathrm{meV}$ for electron-like state and $U_0=-60~\mathrm{meV}$ for hole-like state. Also we chose $\Delta=60~\mathrm{meV}$,  $v{_{_F}}=10^4~\mathrm{cm/s}$, $E_D=10^6 ~\mathrm{V/cm}$, $y_0=100~\mathrm{nm}$, $\ell_0=150~\mathrm{nm}$, $m=0.007m_e$ and $a=5~\mathrm{nm}$.
}
\end{figure}
\begin{figure}
\includegraphics[width=8.5cm,height=4cm]{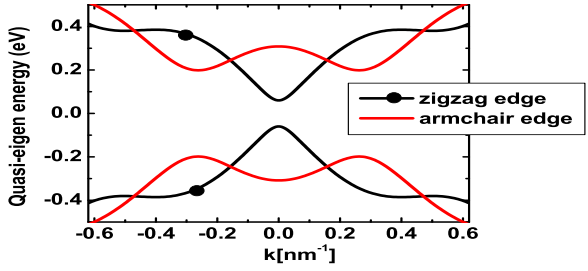}
\caption{\label{fig3}
(Color online) Quasi-eigen energy of GNSLs with zigzag and armchair edges. The material parameters are chosen to be the same as in Fig.~\ref{fig2} but $E_D=10^5 ~\mathrm{V/cm}$ .
}
\end{figure}
\begin{figure}
\includegraphics[width=8.5cm,height=5cm]{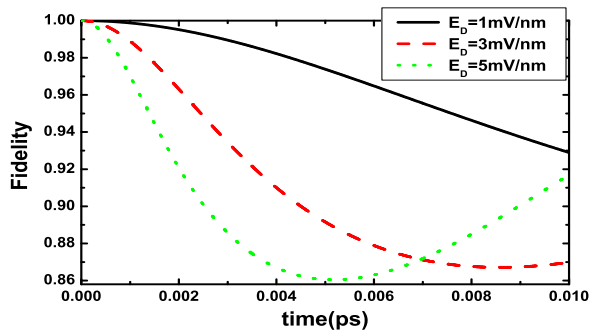}
\caption{\label{fig4}
(Color online) Fidelity, $F=|\langle \chi_-|\chi\left(t\right)\rangle|$,  vs time in GNSLs with zigzag edge. Here we chose $y_0=100~\mathrm{nm}$,  $a=3~\mathrm{nm}$ and $\Delta=10~\mathrm{meV}$.}
\end{figure}
\begin{figure}
\includegraphics[width=8.5cm,height=4cm]{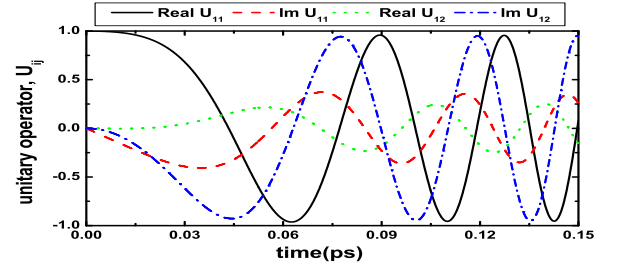}
\caption{\label{fig5}
(Color online) Components of the evolution operator (exact results) for zigzag GNSLs under an applied drift field $E_D=1~\mathrm{mV/nm}$. The material constants are chosen to be the same as in Fig.~\ref{fig4}.}
\end{figure}
\begin{figure}
\includegraphics[width=8.5cm,height=5cm]{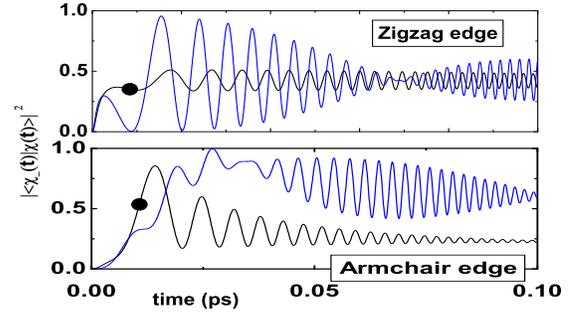}
\caption{\label{fig6}
(Color online) Spin-flip transition probability vs time in both zigzag and armchair GNSLs at $a=1~\mathrm{nm}$ (solid lines with circles) and $a=3~\mathrm{nm}$ (solid lines). The material constants are chosen to be the same as in Fig.~\ref{fig4}. }
\end{figure}
\begin{figure}
\includegraphics[width=8.5cm,height=5cm]{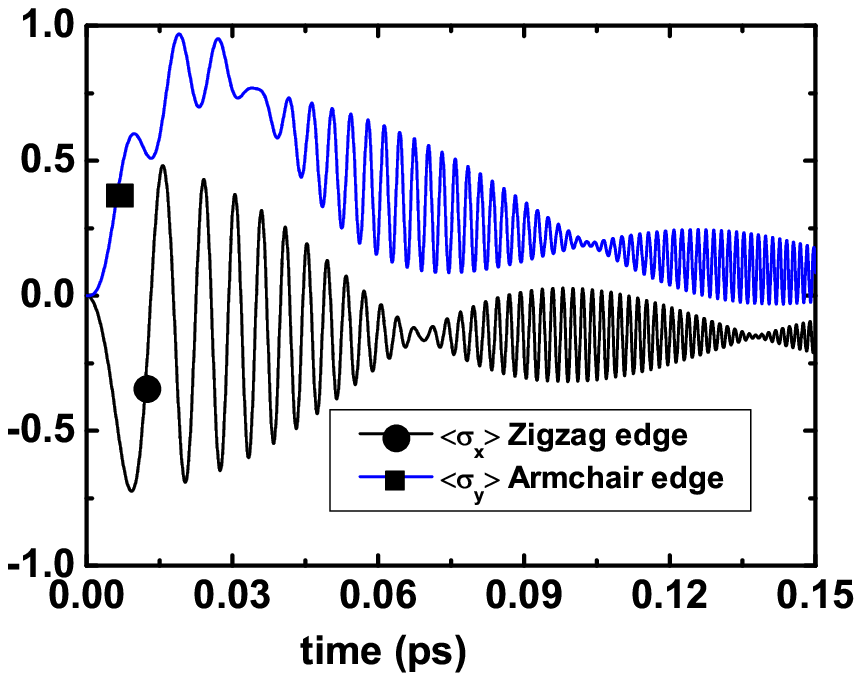}
\caption{\label{fig7}
(Color online) Spin echo is observed in armchair and zigzag GNSLs  under an externally applied drift field. Spin waves propagate in the form of temporal  constructive  and destructive interferences. Here we chose $x_0=y_0=100~\mathrm{nm}$, $E_D=10~\mathrm{mV/nm}$, $a=3~\mathrm{nm}$ and $\Delta=10~\mathrm{meV}$.  }
\end{figure}



The effective mass Hamiltonian for single layer GNSLs  elongated along either armchair or zigzag direction (see Fig.~\ref{fig1})  near the K point of the Brillouin zone can be written as: $ H=v{_{_F}} \boldsymbol {\sigma} \boldsymbol {\cdot} \mathbf{p}  + \Delta\sigma_z + U(r,t)$.
Here $v{_{_F}}$ is the Fermi velocity, $\Delta$ is the miniband width,  $\sigma_i (i=x,y,z)$ are the Pauli spin matrices and $U\left(r,t\right)$ is the confining potential.

For GNSLs with zigzag edge, we consider $ U(r,t)=U_0\left(y+y(t)\right)^2/\ell_0^2$ and  the wave function of the above Hamiltonian $H$ is  $\psi\left(x,y\right)=\exp{\left(i k_x x \right) \phi\left(y\right)}$  (see Ref.~\onlinecite{neto09} for details). Now, we define the relative coordinate $Y=y-y(t)$ and relative momentum $P_Y=p_y-p_y\left(t\right)$ and formulate the total Hamiltonian in the form: $H=H\left(P_Y,Y\right)+H_{qz}$. Here,
\begin{eqnarray}
H\left(P_Y,Y\right)=v{_{_F}}\sigma_yP_Y +U_0 Y^2/\ell_0^2, \label{HPy}\\
H_{qz}= C_1\sigma_x + C_2 \sin\left(k_x a\right)\sigma_y +  \Delta \sigma_z,\label{Hz-quasi}
\end{eqnarray}
where $C_1=\hbar v{_{_F}} k_x$ and $C_2=v{_{_F}} m y_0 \alpha_0 a$. Under an externally applied drift field $E_D$, we write
the equation of motion for $k_x$ as: $\hbar \partial_t k_x(t)=eE_D$ (see Ref.~\onlinecite{krueckl12} for details). Thus we can write $k_x(t)=\alpha_0 t$ with $\alpha_0=eE_D/\hbar$.  In~(\ref{Hz-quasi}), we have written the  quasi momentum  $p_y(t)= m \dot{y}=-my_0\alpha_0 a \sin\left(k_x(t) a\right)$. We have used the Finite Element Method and solved the corresponding eigenvalue problems for~(\ref{HPy}) and~(\ref{Hz-quasi}) and plotted the band structures of GNSLs with zigzag edge in Fig.~\ref{fig2}(a). In addition to the localized states (Fig.~\ref{fig2} (b)), we also see the presence of edge states (Fig.~\ref{fig2} (c)) in GNSLs.

The  Hamiltonian  $H\left(P_Y,Y\right)$ associated to the relative coordinates and relative momentum does not couple to the lowest spin states. Only the quasi-Hamiltonian $H_{qz}$ induces Bloch oscillations in the evolution of spin dynamics in GNSLs.~\cite{krueckl12}  For convenience, we write Eq.~(\ref{Hz-quasi}) as:
\begin{equation}
H_{qz}(t)=C s_+ + C^\star s_- + 2 \Delta s_z, \label{Hz-quasi-1}
\end{equation}
where $C=C_1-iC_2\sin\left(k_x(t) a\right)  $,  $C^\star=conj(C)$ and  $s_{\pm}=s_x\pm is_y$. The band structures of quasi  electron-hole states described by the Hamiltonian~(\ref{Hz-quasi-1}) can be written as
\begin{equation}
\delta=\pm \left[ {|C|}^2  + \Delta^2\right]^{1/2}.\label{Epm}
\end{equation}
The band structure of electron and hole states in the  first Brillouin zone $[-\pi/a,\pi/a]$ with the specific choice of the parameters is shown in Fig.~\ref{fig3}.

We construct a normalized orthogonal set of eigenspinors of the quasi-Hamiltonian~(\ref{Hz-quasi-1}) as:
\begin{equation}
\mathbf{\chi^z_+}\left(t\right)=\frac{\delta-\Delta}{\left[ {{|C|}^2+\left(\delta-\Delta\right)^2}\right]^{1/2}}
\left(\begin{array}{c}
\frac{C}{\delta-\Delta} \\
1\\
\end{array}\right),
\label{chi-p}
\end{equation}
\begin{equation}
\mathbf{\chi^z_-}\left(t\right)=\frac{\delta-\Delta}{\left[ {{|C|}^2+\left(\delta-\Delta\right)^2}\right]^{1/2}}
\left(\begin{array}{c}
-1 \\
\frac{C^\star}{\delta-\Delta} \\
\end{array}\right).
\label{chi-m}
\end{equation}
The above spinors~(\ref{chi-p}) and~(\ref{chi-m}) satisfy $\langle \chi^z_+|\chi^z_+\rangle=\langle \chi^z_-|\chi^z_-\rangle=1$ and $\langle \chi^z_+|\chi^z_-\rangle=\langle \chi^z_-|\chi^z_+\rangle=0$.

With the use of the Feynman disentangling technique, the exact evolution operator of Hamiltonian~(\ref{Hz-quasi-1}) can be written as (see also supplementary material~\onlinecite{supplementary}):
\begin{eqnarray}
U(t,0)&=&T \exp\left\{-\frac{i}{\hbar}\int H(t) dt\right\}\nonumber\\
&=& \exp\left({\alpha(t) s_+}\right) \exp\left({\beta(t) s_0}\right)\exp\left({\gamma(t) s_-}\right),\label{U-2}
\end{eqnarray}
where $T$ is the time ordering operator. At present, the time dependent functions $\alpha,\beta,\gamma$ are unknown and can be found by the Feynman disentangling scheme as discussed below.~\cite{feynman51,popov07,prabhakar10,prabhakar13}

For a spin-1/2 particle, the exact evolution operator~(\ref{U-2}) can be written as:
\begin{equation}
U(t)=\left(\begin{array}{cc}\exp\left\{{\frac{\beta}{2}}\right\} + \alpha\gamma\exp\left\{{-\frac{\beta}{2}}\right\} & ~~~\alpha\exp\left\{-{\frac{\beta}{2}}\right\}\\
\gamma\exp\left\{-{\frac{\beta}{2}}\right\} & ~~~\exp\left\{-{\frac{\beta}{2}}\right\} \end{array}\right), \label{U}
\end{equation}
which can be seen to satisfy $U_{22}=\mathrm{conj}(U_{11})$ and $U_{12}=-\mathrm{conj}(U_{21})$ due to the fact that $U(t)$ is unitary.

At $t=0$, we use the initial condition $\chi^z\left(0\right)=\left(0~1\right)^\top$, where $\top$ denotes transpose and write $\chi^z\left(t\right)=U(t,0)\chi^z\left(0\right)$ as
\begin{equation}
\chi^z\left(t\right)=\left( \alpha\exp\left\{-\beta/2\right\}~~ \exp\left\{-\beta/2\right\}  \right)^\top. \label{chi}
\end{equation}

Next, we find the functional form of $\alpha(t)$, $\beta(t)$,  and $\gamma(t)$ of GNSLs with zigzag edge  by utilizing the Feynman disentangling scheme.~\cite{feynman51,popov07,prabhakar10,prabhakar13}

The exact evolution operator of Hamiltonian~(\ref{Hz-quasi-1})  can be written as:
\begin{eqnarray}
U(t)
&=& \exp{\left\{\alpha (t) s_+\right\}}T  \nonumber\\
&&\exp\left[-\frac{i}{\hbar}\int_0^t  \left\{ \left(C-x  \right)s'_+ + C^\star s'_- + 2 \Delta s'_0  \right\} dt'\right],~~~~~~ \label{U-3a}
\end{eqnarray}
where
\begin{eqnarray}
\alpha\left(t\right)=-\frac{i}{\hbar}\int^t_0 x(t) dt',\label{alpha}\\
s'_\mu\left(t\right)=\exp\left(-\alpha s_+ \right) s_\mu \exp\left(\alpha s_+ \right).
\label{s-mu-prim}
\end{eqnarray}
Differentiating Eq.~(\ref{s-mu-prim}) with respect to $\alpha$, we find
\begin{equation}
\frac{d s'_\mu\left(t\right)}{d\alpha}=-\exp\left(-\alpha s_+ \right) \left(s_+s_\mu-s_\mu s_+\right) \exp\left(\alpha s_+ \right). \label{ds}
\end{equation}
By utilizing initial condition $s'_\mu\left(0\right)=s_\mu$, the primed operators are determined
\begin{equation}
s'_+=s_+,~s'_0=s_+\alpha+s_0,~s'_-=s_--s_+\alpha^2-2s_0\alpha.
\label{s-mu-prim-1}
\end{equation}
By substituting Eq.~(\ref{s-mu-prim-1}) in Eq.~(\ref{U-3a}) and equating the coefficient of $s_+$ to zero, we find the following Riccatti equation:
\begin{equation}
\frac{d \alpha}{dt}= -\frac{i}{\hbar}\left\{C-C^\star \alpha^2 + 2\Delta \alpha \right\}. \label{alpha-1}
\end{equation}
Hence the dependence on $s_+$ in~(\ref{U-3a}) has been disentangled. By following the above procedure and by disentangling $s_0$ and $s_-$, another set of the Riccatti equations for the time dependent functions $\beta(t)$ and $\gamma(t)$ can be written as
\begin{eqnarray}
\frac{d \beta}{dt}= -\frac{2i}{\hbar}\left\{\Delta -C^\star \alpha   \right\}, \label{beta-1}\\
\frac{d \gamma}{dt}= -\frac{i}{\hbar}C^\star \exp\left\{\beta\right\}. \label{gamma-1}
\end{eqnarray}

In a similar way, for armchair GNSLs, we write the total Hamiltonian $H=H\left(P_X,X\right)+H_{qa}$, where
\begin{eqnarray}
H\left(P_X,X\right)=v{_{_F}}\sigma_xP_x +U_0 X^2/\ell_0^2, \label{HPx}\\
H_{qa}= -C'_2\cos\left(k_y(t) a\right) \sigma_x + C'_1 \sigma_y + \Delta \sigma_z  .\label{Ha-quasi}
\end{eqnarray}
Here $C_1'=\hbar v{_{_F}} k_y(t)$ and $C_2'=v{_{_F}} m x_0 \alpha_0 a$.
We construct a normalized orthogonal set of eigenspinors of the quasi-Hamiltonian~(\ref{Ha-quasi}) as:
\begin{equation}
\mathbf{\chi^a_+}\left(t\right)=\frac{\delta-\Delta}{\left[ {{|C'|}^2+\left(\delta-\Delta\right)^2} \right]^{1/2}}
\left(\begin{array}{c}
-1\\
\frac{-i{C'}^\star}{\Delta-\delta} \\
\end{array}\right),
\label{chi-pa}
\end{equation}
\begin{equation}
\mathbf{\chi^a_-}\left(t\right)=\frac{\delta-\Delta}{\left[ {{|C'|}^2+\left(\delta-\Delta\right)^2}\right]^{1/2}}
\left(\begin{array}{c}
\frac{iC'}{\Delta-\delta}  \\
1\\
\end{array}\right),
\label{chi-ma}
\end{equation}
where $C'=C'_1-iC'_2\cos\left(a \alpha_0 t \right)$.
At $t=0$, we use the initial condition
\begin{equation}
\mathbf{\chi^a}\left(0\right)=\frac{\delta-\Delta}{\left[C_2'^2+\left(\delta-\Delta\right)^2\right]^{1/2}}
\left(\begin{array}{c}
\frac{C'_2}{ \Delta-\delta}  \\
1\\
\end{array}\right),
\label{chia}
\end{equation}
and find $\chi^a\left(t\right)=U(t,0)\chi^a\left(0\right)$.
The three coupled Riccatti equations for the quasi-Hamiltonian of  GNSLs with armchair edge can be written as
\begin{eqnarray}
\frac{d \alpha}{dt}= -\frac{i}{\hbar}\left\{\tilde{C}-{\tilde{C}}^\star \alpha^2 + 2\Delta \alpha \right\}, \label{alpha-2}\\
\frac{d \beta}{dt}= -\frac{2i}{\hbar}\left\{\Delta -{\tilde{C}}^\star \alpha   \right\}, \label{beta-2}\\
\frac{d \gamma}{dt}= -\frac{i}{\hbar}{\tilde{C}}^\star \exp\left\{\beta\right\}, \label{gamma-2}
\end{eqnarray}
where $\tilde{C}=-iC'$. We have solved the three coupled Riccatti equations for zigzag and armchair GNSLs numerically and found the exact evolution operator~(\ref{U}).
The supplementary material provides an  additional example of finding exact evolution operator for a spin-1/2 particle in presence of tilted magnetic field.~\cite{supplementary}


In Fig.~\ref{fig4}, we have plotted the fidelity vs time for several different values of the applied drift field. As expected from the analytical solutions (see Eqs.~\ref{chi-m} and~\ref{chi}), with the increasing drift field, the fidelity becomes enhanced because the amplitude of the spin polarization increases as the magnitude of the applied drift field is increased while the spatial frequency of the propagating waves is reduced.  These effects cause dips in fidelity at an early stage with increasing magnitude of the drift field.

In Fig.~\ref{fig5}, we study the evolution operator in graphene under a drift field. The components of $U_{ij}$ are superposed leading to constructive and destructive interferences. We observe   Hahn echo patterns in a constructive interference regime of  the evolution of spin dynamics.\cite{belonenko12,hahn90} This will be separately discussed with reference to Fig.~\ref{fig7}. Spin flip occurs when the localized states in GNSLs with armchair and zigzag edges move rapidly along the ribbon with the application of the time dependent gate controlled electric field. Such spin flip probabilities are plotted in Fig.~\ref{fig6} for both the armchair and zigzag GNSLs at different values of the period of the superlattices. As can be seen, the spin-flip probabilities in both the armchair and zigzag GNSLs are enhanced with  increasing value of the period of the superlattice. This is  due to the fact that the  enhancement in the amplitude of the spin waves occur with the increasing value of the period of the superlattice.

In Fig.~\ref{fig7}, we have plotted the expectation values of the Pauli spin matrices  vs time  under an externally applied drift field in both GNSLs with zigzag and armchair edges.  For $t>0$, the components of the spin waves  fluctuate rapidly and thus we observed Hahn echo patterns  in GNSLs. In this regime, where the spin echo can be observed, the spin waves  are  superposed which  induce constructive interference. When the superposed spin waves induce destructive interference, we find strong beating patterns. The Hahn echo accompanied by strong beating patterns can be seen at different intervals of time in GNSLs with zigzag (solid line with circles) and armchair (solid lines with diamond) edges due to the fact that the spin waves in both kinds of superlattices travel with different velocities. Note that the  Hahn echo patterns in $\langle\sigma_x\rangle$ of zigzag edge resemble to the $\langle\sigma_y\rangle$  of armchair edge. This is due to the fact that the coefficients of $\sigma_x$ and $\sigma_y$ of the quasi-Hamiltonian of the GNSLs with zigzag and armchair edges Eq.~(\ref{Hz-quasi}) and~(\ref{Ha-quasi}) respectively interchanges during the evolution of spin dynamics.

To conclude, we have shown that the quasi-Hamiltonian of the GNSLs with zigzag and armchair edges can be used to investigate the evolution of spin dynamics under an externally applied drift field. We have shown that the exact evolution operator can be found for such Hamiltonian  via the Feynman disentangling technique. With a particular choice of the period and drift field, the spin polarization can  be controlled to propagate  on the surface of the Bloch sphere in a desired fashion. During the transportation of the spin on the surface of the Bloch sphere, the Hahn echo, accompanied by strong beating patterns, is observed.
The authors in Ref.~\onlinecite{song13} have  fabricated a field effect transistors (FET) using graphene where the gate electrode was placed over a thin polymer gate dielectric layer. Such devices show excellent output and transfer characteristics (mobility $\approx 0.1 \mathrm{cm^2/Vs}$) for gate and drain operations (see Fig.(4b) in Ref.~\onlinecite{song13}). The present work was motivated in part by these experimental studies and our results (e.g., Figs.~\ref{fig1}-\ref{fig7}) might provide useful information for the design of quantum logic gates based on the application of the externally applied drift fields in GNSLs with both the armchair and zigzag edges.

This work has been supported by NSERC (Canada) and Canada Research Chair programs.

\section*{Supplementary materials:}

To verify  that the evolution operator~(Eq.~\ref{U}) is exact, we provide an example associated with the Hamiltonian of spin-1/2 particle in an effective magnetic
field having a known solution:~\cite{grifith_book}
\begin{equation}
H(t)=\Omega \cos\left(\omega t\right) s_x+\Omega \sin\left(\omega t\right) s_y+\Delta s_z.
\label{Ht-2}
\end{equation}
The energy eigenvalues of~(\ref{Ht-2}) are $\varepsilon_{\pm}=\pm \hbar\delta /2,$ where $\delta=\left[\Delta^2+\Omega^2\right]^{1/2}$.
We construct a normalized orthogonal set of eigenspinors of Hamiltonian~(\ref{Ht-2}) as:
\begin{eqnarray}
\mathbf{\chi_+}\left(t\right)=\frac{1}{\sqrt N}
\left(\begin{array}{c}
\Omega \\
\left(\delta - \Delta\right)\exp\left\{i\omega t\right\} \\
\end{array}\right),\label{chi-p-1}\\
~\nonumber\\
\mathbf{\chi_-}\left(t\right)=\frac{1}{\sqrt N}
\left(\begin{array}{c}
\left(\delta - \Delta\right)\exp\left\{-i\omega t\right\} \\
-\Omega \\
\end{array}\right),
\label{chi-m-1}
\end{eqnarray}
where
\begin{equation}
N^2=\Omega^2+\left(\delta-\Delta\right)^2.
\label{N-1}
\end{equation}
Since the Hamiltonian~(\ref{Ht-2}) is time dependent, the general time dependent Schr\"{o}dinger equations can be written as
\begin{eqnarray}
\partial_t a(t)=-\frac{i}{2}\left(\Delta a + \Omega \exp\left\{-i \omega t\right\}b\right),\label{pdt-1}\\
\partial_t b(t)=-\frac{i}{2}\left( \Omega \exp\left\{i \omega t\right\}a - \Delta b\right).\label{pdt-2}
\end{eqnarray}
The exact  solution of time dependent Schr\"{o}dinger Eqs.~(\ref{pdt-1}) and~(\ref{pdt-2}) can be written as:~\cite{grifith_book}
\begin{widetext}
\begin{eqnarray}
a\left( t \right)= \frac{\Omega}{\sqrt N}\left\{ \cos\left( \frac{\lambda' t}{2} \right) -\frac{i}{\lambda'}\left(\delta-\omega\right) \sin\left( \frac{\lambda' t}{2} \right)\right\}\exp\left(-\frac{i\omega t}{2}\right),  \label{chi-p-2}\\
b\left( t \right)= \frac{\delta-\Delta}{\sqrt N}  \left\{ \cos\left( \frac{\lambda' t}{2} \right) -\frac{i}{\lambda'}\left(\delta+\omega\right) \sin\left( \frac{\lambda' t}{2} \right)\right\}\exp\left(\frac{i\omega t}{2}\right), \label{chi-m-2}
\end{eqnarray}
\end{widetext}
where
\begin{equation}
\lambda'=\sqrt {\Omega^2+\left(\Delta-\omega\right)^2}. \label{lambda-p}
\end{equation}
Expressing~(\ref{chi-p-2}) and~(\ref{chi-m-2}) as a linear combination of $|\chi_+\rangle$ and $|\chi_-\rangle$, we have
\begin{widetext}
\begin{equation}
\psi\left( t \right)=\left\{ \cos\left( \frac{\lambda' t}{2} \right) -\frac{i}{\lambda'}\left(\delta-\frac{\omega \Delta}{\delta}\right) \sin\left( \frac{\lambda' t}{2} \right)\right\}
\exp\left(-\frac{i\omega t}{2}\right)\chi_+\left( t \right)+i\frac{\omega\Omega}{\lambda'\delta}\sin\left( \frac{\lambda' t}{2} \right)
\exp\left(\frac{i\omega t}{2}\right)\chi_-\left( t \right).\label{chi-1}
\end{equation}
\end{widetext}
Clearly, one can immediately write  the transition probability:
\begin{equation}
|\langle\chi_-\left( t \right)|\psi\left( t \right)\rangle|^2=\left(\frac{\omega\Omega}{\lambda'\delta}\right)^2\sin^2\left( \frac{\lambda' t}{2} \right), \label{probability-1}\\
\end{equation}
provided that
\begin{equation}
|\langle\chi_-\left( t \right)|\psi\left( t \right)\rangle|^2+|\langle\chi_+\left( t \right)|\chi\left( t \right)\rangle|^2=1.\label{probability-2}
\end{equation}
\begin{figure}
\includegraphics[width=8.5cm,height=5cm]{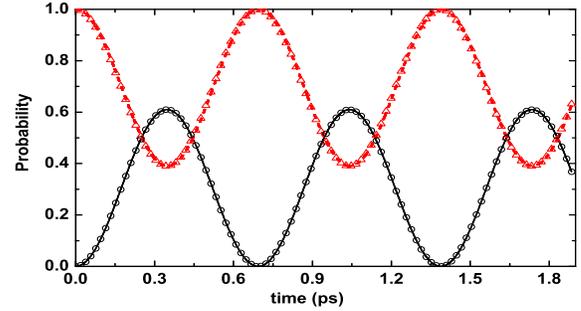}
\caption{\label{fig8}
(Color online) Transition probability vs time. Here we chose $\omega=10~\mathrm{THz}$ and $\Omega=\Delta=1~\mathrm{THz}$. Transition probabilities  obtained from Eqs.~(\ref{probability-1}) and~(\ref{probability-2}) (solid and dashed lines, respectively) are seen to be in excellent agreement to the ones obtained from the Feynman disentangling operator scheme. The transition probabilities are given by $|\langle \chi_+\left(t\right)|U\left(t,0\right) \chi \left(0\right)\rangle|^2$ (circles) and $|\langle \chi_-\left(t\right)|U\left(t,0\right) \chi \left(0\right)\rangle|^2$ (triangles).
}
\end{figure}
Next, we apply the Feynman disentangling operator scheme and find the exact evolution operator of the Hamiltonian~(\ref{Ht-2}) which initially takes the form of~(Eq.9, see the paper) but with three different coupled Riccati equations (evolved during the disentangling operator scheme of the Hamiltonian~(\ref{Ht-2})). These are:
\begin{eqnarray}
\frac{d \alpha}{dt}= -\frac{i}{\hbar}\left\{  \frac{\Omega}{2} \exp\left(-i\omega t\right) +\Delta \alpha - \frac{\Omega}{2} \exp\left(i\omega t\right) \alpha^2  \right\},~~~~ \label{alpha-2}\\
\frac{d \beta}{dt}= -\frac{i}{\hbar}\left\{\Delta-\Omega \exp\left(i\omega t\right) \alpha   \right\},~~~~~~ \label{beta-2}\\
\frac{d \gamma}{dt}= -\frac{i}{\hbar}\frac{\Omega}{2}\exp\left\{\beta+i\omega t\right\}.~~~~~~ \label{gamma-2}
\end{eqnarray}
Usually, an exact solution of such Riccati  differential equations does not exist. However, in this case, it is possible to find the exact solution of  Eqs.~(\ref{alpha-2}),~(\ref{beta-2}) and~(\ref{gamma-2}) as:
\begin{eqnarray}
\alpha\left(t\right)=\frac{\exp\left(-i\lambda t\right)-\exp\left(-i\omega t\right)}{n_2-n_1 \exp\left(-i\varpi t\right)},\label{alpha-3}\\
\beta\left(t\right)=\frac{\exp\left(-i\lambda t\right)\left(n_2-n_1\right)^2}{\left\{n_2-n_1 \exp\left(-i\varpi t\right)\right\}^2},\label{beta-3}\\
\gamma\left(t\right)=\frac{\exp\left(-i\varpi t\right)-1}{n_2-n_1 \exp\left(-i\varpi t\right)},\label{gamma-3}
\end{eqnarray}
where $\lambda=\varpi+\omega$,  $\varpi=\Omega\left(n_2-n_1\right)/2$,
\begin{eqnarray}
n_1= \frac{\Delta-\omega}{\Omega}+\frac{\left[\left(\Delta-\omega\right)^2+\Omega^2\right]^{1/2}}{\Omega},\label{n1}   \\
n_2= \frac{\Delta-\omega}{\Omega}-\frac{\left[\left(\Delta-\omega\right)^2+\Omega^2\right]^{1/2}}{\Omega}.\label{n2}
\end{eqnarray}
In Fig.~\ref{fig8}, the transition probability  obtained from Eqs.~(\ref{probability-1}) and (\ref{probability-2}) (solid and dashed lines) is seen to be in excellent agreement to the one obtained from the disentangling  scheme (circles and triangles). Thus, we have demonstrated  that the evolution operator obtained from the disentangling operator scheme is exact. Finding an exact unitary operator is one of the requirements for quantum computing and is one of the motivations of the present work.


\begin{thebibliography}{35}
\bibitem [{sup()}]{supplementary}%
  \BibitemOpen
  \href@noop {} {\bibinfo  {journal} {See supplentary material at [URL will be
  inserted by AIP] for spin-1/2 particle in a tilted magnetic field}\
  }\BibitemShut {NoStop}%
\end{thebibliography}

%

\end{document}